\documentclass[aps,pra,a4paper,12pt,tightenlines,superscriptaddress,notitlepage]{revtex4-1}
\makeatletter\newcommand{\@ptsize}{2}\makeatother
\usepackage{typearea}
\usepackage[american]{babel}
\usepackage[T1]{fontenc}
\usepackage[utf8]{inputenc}
\usepackage{amsmath}
\usepackage{txfonts}\renewcommand{\pi}{\piup}\newcommand{\Deltaup}{\Delta}
\usepackage{textcomp}
\usepackage{microtype}
\usepackage{graphicx}
\usepackage[product-units=power,retain-explicit-plus,per-mode=power,per-symbol=$/$]{siunitx}
\usepackage{hyperref}
\newcommand{\gsf}{1.375}
\renewcommand{\vec}{\boldsymbol}
\newcommand{\TUKL}{Department of Physics and Research Center OPTIMAS, University of Kaiserslautern, 67663~Kaiserslautern, Germany}
\newcommand{\IPM}{Fraunhofer Institute for Physical Measurement Techniques IPM, 79110~Freiburg, Germany}
\newcommand{\EIT}{Department of Electrical and Computer Engineering and Research Center OPTIMAS, University of Kaiserslautern, 67663~Kaiserlautern, Germany}

\begin{document}

\title{\Large\sffamily In-Plane Focusing of Terahertz Surface Waves on a Gradient Index Metamaterial Film}
\date{\today}
\author{Martin~F.~Volk}\affiliation{\TUKL}\affiliation{\IPM}
\author{Benjamin~Reinhard}\email{breinhard@physik.uni-kl.de}\affiliation{\TUKL}\affiliation{\EIT}
\author{Jens~Neu}\affiliation{\TUKL}\affiliation{\EIT}
\author{René~Beigang}\affiliation{\TUKL}\affiliation{\IPM}
\author{Marco~Rahm}\affiliation{\EIT}
\begin{abstract}
\noindent
We designed and implemented a gradient index metasurface for the in-plane focusing of confined terahertz surface waves.
We measured the spatial propagation of the surface waves by two-dimensional mapping of the complex electric field using a terahertz near-field spectroscope.
The surface waves were focused to a diameter of \SI{500}{\um} after a focal length of approx.\ \SI{2}{\mm}.
In the focus, we measured a field amplitude enhancement of a factor of \num{3}.
\end{abstract}
\maketitle

It has recently been demonstrated that the sensitivity, resolution, and signal-to-noise ratio of sensing devices operating in the terahertz (THz) frequency range (approx.\ \SIrange{0.1}{10}{\THz}) can be significantly improved by exploiting the strongly localized fields at the resonant elements of electromagnetic metamaterials \cite{driscoll2007,ohara2008,tao2010,reinhard2012b}.
Another approach to improve the performance of THz sensors is to make use of propagating guided waves in order to increase the interaction length between the THz field and the sample \cite{rau2005,isaac2008,theuer2010}.
In the near infrared and visible regions of the electromagnetic spectrum, surface plasmon polaritons (SPPs) at the interface between a metal and a dielectric offer a convenient way to control propagating localized waves at very small length scales \cite{barnes2003}.
At THz frequencies, however, where the conductivity of most metals is very high, SPPs are only very weakly localized, which limits their usefulness considerably.
A promising way to increase the localization is to use metamaterial surfaces composed of arrays of sub-wavelength sized resonator elements \cite{navarro-cia2009,reinhard2010,minowa2011}.
These offer a great flexibility to deliberately design the electromagnetic response of the metasurface at the unit cell level and thus to gain control over the spatial propagation properties of the confined THz surface waves.
Based on this general concept, it is possible to devise a new class of integrated optical systems for THz waves that might pave the way towards innovative, compact and budget-priced THz measurement systems.

Here, we designed and fabricated a gradient index (GRIN) metamaterial film that focuses strongly confined surface waves within the plane of propagation.
We verified the focusing capabilities by measuring the complex electric field amplitude of the surface waves using a THz near-field spectroscope.
The metasurface consisted of split-ring resonators (SRRs) with spatially varying geometric parameters.
As previously shown \cite{reinhard2010}, an SRR-based metasurface supports confined waves with transverse-electric (TE) polarization, which means that the magnetic field orientation is normal to the surface.

\begin{figure}%
\centering%
\includegraphics[scale=\gsf]{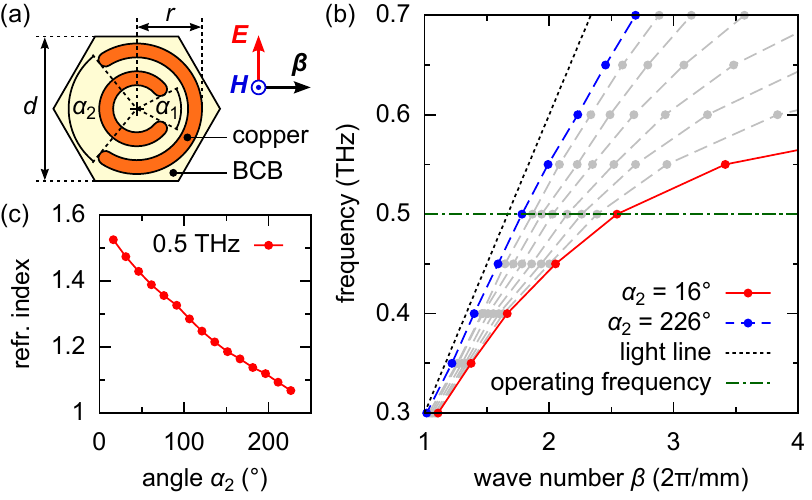}%
\caption{(a) Unit cell of the metamaterial: $d=\SI{52}{\um}$, $r=\SI{23.5}{\um}$, $\alpha_1=\ang{53.7}$.
The width of the metal strips and the gap between the inner and outer ring are equal to \SI{5}{\um}.
The opening angle $\alpha_2$ is varied to change the refractive index.
Vectors $\vec E$, $\vec H$, and $\vec\beta$ indicate the direction of the electric field, magnetic field, and wave vector of the localized waves, respectively.
(b) Dispersion relation of the surface waves for different values of the opening angle $\alpha_2$ (\ang{16} to \ang{226} in steps of \ang{30}).
(c) Dependence of the refractive index of the metamaterial structure on the opening angle $\alpha_2$ at a frequency of \SI{0.5}{THz}.
The data in (b) and (c) is taken from numerical calculations performed with \emph{CST Microwave Studio\textsuperscript\textregistered} simulation software.}%
\label{fig:refractive_index}%
\end{figure}

As a first step, we designed the unit cells of the metamaterial film which consisted of a single layer of hexagonally ordered double SRRs embedded in a \SI{20}{\um} thick dielectric matrix of benzocyclobutene (BCB). 
The geometry parameters of a typical unit cell are shown in Fig.~\ref{fig:refractive_index}(a).
We tuned the resonance frequency and thus the dispersion of the refractive index by variation of the opening angle $\alpha_2$ of the outer ring, which changes the capacitance of the SRR.
This is shown in Fig.~\ref{fig:refractive_index}(b) where we plotted the numerically calculated dispersion relation of the surface waves for different values of $\alpha_2$.
The corresponding effective refractive index of the metasurface can be defined as $n = c_0\beta/\omega$, where $c_0$ is the speed of light in vacuum, $\beta$ is the wave number, and $\omega$ is the angular frequency of the surface wave.
For localized waves, $n$ is larger than the refractive index of the surrounding media.
Figure \ref{fig:refractive_index}(c) illustrates the dependence of the refractive index on the opening angle $\alpha_2$ at an operating frequency of \SI{.5}{\THz}.
By changing the angle $\alpha_2$ we could tune the refractive index between \num{1.07} and \num{1.53}.

\begin{figure}%
\centering%
\includegraphics[scale=\gsf]{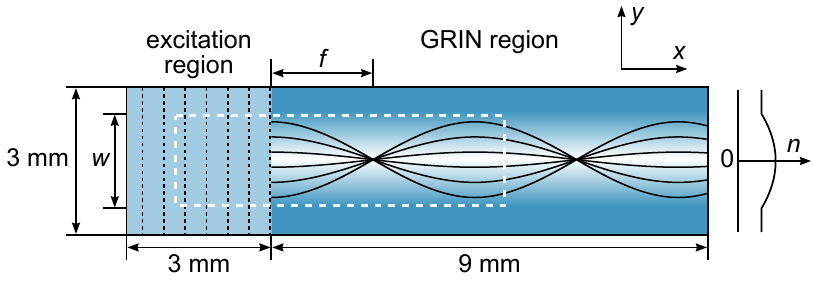}%
\caption{Layout of the GRIN metasurface with an artificial grating in the excitation region and a parabolic refractive index profile in the GRIN region where the confined surface waves propagate and are focused.
The refractive index $n$ varies between \num{1.5} (center line) and \num{1.1} (edge lines).
The focal length is $f = \SI{2.15}{\mm}$.
The dashed white line denotes the region that we mapped in the near-field measurement.}%
\label{fig:layout}%
\end{figure}

In the next step, we exploited the tunability of the refractive index to design a GRIN metamaterial film that periodically focuses plane surface waves at a focal length $f$ in the plane of propagation.
For this purpose, we spatially arranged unit cells of different refractive index to obtain a parabolic refractive index profile normal to the propagation direction $x$ of the surface wave \cite{hecht2002}. The spatially dependent refractive index can be described by
\begin{equation}
n(y) = n_\mathrm{c} - \Deltaup n \frac{4y^2}{w^2}~,
\end{equation}%
where $n_\mathrm{c}$ is the refractive index along the line in $x$ direction which is centered with respect to the width $w$ of the GRIN structure in $y$ direction, $\Deltaup n$ is the refractive index difference between the values at the center line and the edge lines, and $y$ is the distance from the center line. 
The focal length $f$ is then given by
\begin{equation}
f = \frac{\pi}{4}w\sqrt{\frac{n_\mathrm{c}}{2\Deltaup n}}~.
\end{equation}
For our design, we chose $n_\mathrm{c} = \num{1.5}$, $\Deltaup n = \num{0.4}$, and $w = \SI{2}{\mm}$. This results in a focal length $f = \SI{2.15}{\mm}$.
Figure~\ref{fig:layout} schematically shows the layout of the metamaterial.
At the edges of the structure in $y$ direction, we added \SI{0.5}{\mm} wide strips with constant refractive index $n=n_\mathrm{c}-\Deltaup n = \num{1.1}$ in order to minimize boundary effects.
Thus, the total width of the metamaterial film was \SI{3}{\mm}.
Because of the anisotropic response of the SRRs, the propagation properties of the bound waves are strongly dependent on the orientation of the electric field with respect to the gaps of the SRRs.
To ease the design procedure, we rotated each of the SRRs to align them with the local electric field and wave vector of the surface waves as indicated in Fig.~\ref{fig:refractive_index}(a).
For the excitation of the surface waves we coupled freely propagating THz waves into a defined bound surface wave mode.
For this purpose we defined an excitation zone where we implemented an artificial grating by leaving out every 10th column of SRRs in $x$ direction.
That way we obtained a grating period of \SI{450}{\um}.
Within the excitation region, the refractive index was constant ($n = \num{1.33}$).

For the fabrication of the metamaterial films we employed standard micro- and nanofabrication methods such as UV lithography and spin coating.
We embedded \SI{300}{\nm} thick copper SRRs between two \SI{10}{\um} thick layers of BCB (\emph{Cyclotene\textsuperscript\textregistered 3022-63} resin from \emph{The Dow Chemical Company}), resulting in flexible metamaterial films of \SI{20}{\um} thickness.
Details on the general fabrication process can be found in a previous publication \cite{paul2008}.

\begin{figure}%
\centering%
\includegraphics[scale=\gsf]{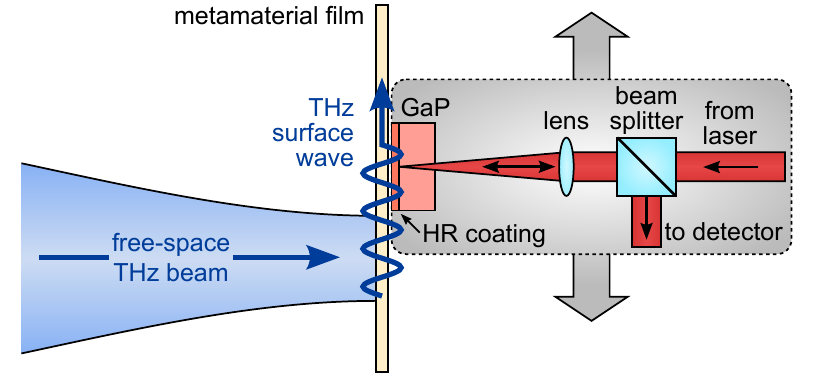}%
\caption{Schematic of the near-field spectroscope.
The GaP detector crystal is brought in direct contact with the metamaterial.
The detector unit can be moved in two dimensions relative to the sample, while the position of the THz beam is fixed.}%
\label{fig:setup}%
\end{figure}

In order to experimentally confirm the focusing capability of the GRIN metasurface, we measured the spatial distribution of the complex electric field amplitude using a near-field THz time-domain spectroscope (THz-TDS) based on electro-optic sampling (EOS) \cite{valk2002,neu2010a}.
Figure~\ref{fig:setup} shows a schematic of the detection setup.
We generated short, broadband THz pulses by illuminating a photoconductive switch with \SI{40}{fs} long infrared laser pulses from a Ti:sapphire laser.
With a pair of off-axis parabolic mirrors, we focused the THz beam on the excitation region of the metamaterial film.
The grating lines of the excitation region were parallel to the electric field of the THz pulse.
At the opposite side of the metamaterial film, we brought a $\SI{2x2x0.4}{\mm}$ (110)-cut gallium phosphide (GaP) crystal in direct contact with the metamaterial.
A fraction of the laser pulse that excited the THz pulse was used as a probe pulse and focused onto the surface of the GaP crystal that faced the metamaterial.
This surface was coated with a highly reflective (HR) layer for the wavelength of the probe beam while the entrance facet for the probe pulse was coated with an anti-reflection layer.
We then analyzed the reflected probe pulse by a combination of a quarter-wave plate, a Wollaston prism, and a balanced photodetector \cite{neu2010a} to obtain a signal proportional to the electric THz field inside the GaP crystal.
The entire detector unit was mounted on a two-dimensional positioning stage.
By raster scanning the metamaterial film, we obtained a 2-D map of the complex electric field of the confined THz surface waves. 

\begin{figure}%
\centering%
\includegraphics[scale=\gsf]{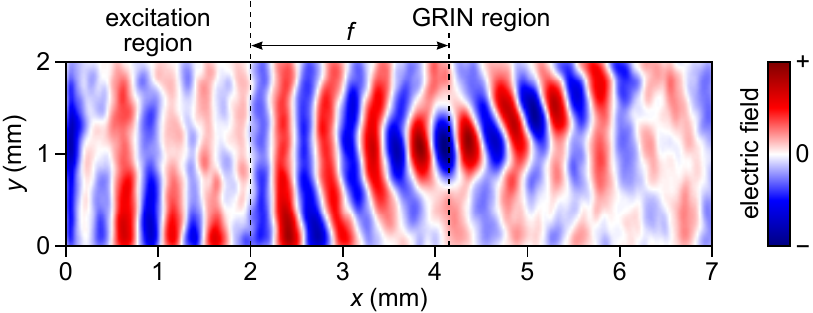}%
\caption{
Measured electric near field of the GRIN metasurface.
The real part of the $y$ component of the complex electric field amplitude at \SI{0.45}{\THz} is shown.
}%
\label{fig:measurements}%
\end{figure}

It should be noted that although we originally designed the GRIN structure to operate at a frequency of \SI{0.5}{\THz}, we obtained the best focusing and the weakest attenuation at a slightly lower frequency (\SI{0.45}{\THz}).
Despite the dispersive response of the SRRs, the focal length of the GRIN metasurface at a frequency of \SI{0.45}{\THz} differs only moderately from the design value so that the focusing ability of the structure is maintained.
Figure~\ref{fig:measurements} shows a measurement of the localized electric field of the surface waves at \SI{0.45}{THz}.
In this case, we raster scanned the near field of the surface waves with a spatial resolution of \SI{100}{\um} in $y$ direction and \SI{50}{\um} in $x$ direction.
At each measurement point, we measured a time trace of \SI{48}{\ps}.
The total number of time traces was \num{20x140}, the total area covered by the measurement was \SI{2x7}{\mm}.
To obtain the 2-D near-field maps shown in Fig.~\ref{fig:measurements}, we transformed each time signal into the frequency domain and selected the complex field amplitude at the desired frequency.
Figure \ref{fig:measurements} shows the real part of the $y$ component of the complex electric field amplitude.
The data was filtered in wave vector space with a gaussian-shaped low-pass filter to reduce noise.
Furthermore, waves with $\left|\beta_x\right| < 1\times\SI{2\pi}{\per\mm}$ were cut out because they correspond to the fraction of the THz beam that is transmitted through the sample without coupling to surface waves.
In the excitation region, at $x < \SI{2}{\mm}$, free space THz waves are diffracted perpendicularly into the metamaterial membrane.
Due to the parabolic index profile, the phase velocity of the surface wave phase fronts is smaller along the center line of the GRIN region than towards the edge lines.
For this reason, the surface waves converge to a \SI{500}{\um} wide focus located approx.\ \SI{2}{\mm} behind the excitation region.
In the focus, we measured a field amplitude enhancement of a factor of \num{3} compared to the field directly behind the grating.

In conclusion, we designed and fabricated a gradient index metamaterial film that focuses confined terahertz surface waves within the plane of propagation.
The metamaterial film was comprised of split-ring resonators, thus supporting transverse-electric surface waves.
By variation of the geometry of the SRRs we created a gradient of the refractive index with a parabolic profile.
The refractive index could be tuned between \num{1.07} and \num{1.53} at an operating frequency of \SI{.5}{\THz}.
By use of near-field spectroscopy we mapped the 2-D distribution of the complex electric THz field and verified the focusing behavior of the GRIN metasurface.
The presented design procedure is very versatile and can be applied to implement further GRIN structures or transformation optics devices for the construction of integrated THz systems.

We gratefully acknowledge support for the sample fabrication from the Nano Structuring Center (NSC) at the University of Kaiserslautern, Germany.
The work was funded by the German Research Foundation (DFG) under Grant No.\ RA~1903/2-2.

\end{document}